\documentclass[12pt,timesmath]{article}

\usepackage{amsmath}
\usepackage{amsthm}
\usepackage{amssymb}
\usepackage{stmaryrd}
\usepackage{graphicx}
\usepackage{latexsym}
\usepackage{setspace}
\usepackage{braket}
\newtheorem{ass}{General Assumption}
\newtheorem{theorem}[ass]{Theorem}
\newtheorem{lemma}[ass]{Lemma}
\newtheorem{definition}[ass]{Definition}

\newtheorem{rem}[ass]{Remark}

\begin{document}

% Uncomment for double-spacing \doublespacing

\title{A new result on the Klein-Gordon equation in the background of a rotating black hole}

\author{Horst R. Beyer \\ Louisiana State University (LSU)\\
Center for Computation \& Technology (CCT)\\
235 Johnston Hall \\
 LA 70803, USA \\
\& \\
 Max Planck Institute for Gravitational Physics\\
 Albert Einstein Institute (AEI)\\
Am M\"uhlenberg 1 \\
 D-14476 Potsdam, Germany}

\date{\today}                                     
% Uncomment if TWO COLUMNS REQUIRED
%\twocolumn

\maketitle

\begin{abstract}
This short paper should serve as basis for further 
analysis of a previously found new symmetry of the solutions of 
the wave equation 
in the gravitational field of a Kerr black hole. Its main 
new result is the proof of essential self-adjointness of the  
spatial part of a reduced normalized wave operator of
the Kerr metric in a weighted $L^2$-space. As a consequence,  
it leads to a purely operator theoretic proof of the 
well-posedness of the initial value problem of the reduced 
Klein-Gordon equation in that field in that $L^2$-space and 
in this way generalizes a corresponding result of Kay ($1985$) 
in the case of the Schwarzschild black hole. It is believed that 
the employed methods are applicable to  
other separable wave equations. 
\end{abstract}

%Uncomment out if preprint format required
%\pacs{00.00, 20.00, 42.10}

\section{General Introduction}

The linearized stability of the Schwarzschild black hole follows by 
combination 
of the Regge-Wheeler-Zerilli-Moncrief decomposition of gravitational 
perturbations of the Schwarz\-schild metric \cite{reggewheeler,zerilli,moncrief} and 
a result  by Kay and Wald \cite{kaywald} that proves the boundedness of 
all solutions of the wave equation corresponding to $C^{\infty}$-data of compact 
support. The proof of the last rests on the positivity of 
the conserved energy. 
\newline  
\linebreak
The question of the linearized stability of the Kerr black hole is 
still an open problem whose outcome is of major importance 
to General Relativity. In comparison to the case of the 
Schwarzschild black hole, the solution to this problem is 
considerably more complicated. Mainly, this is due to two facts.     
First, a decomposition comparable to that of 
Regge-Wheeler-Zerilli-Moncrief does not yet exist in this case, although 
the recent finding of \cite{beyercraciun} gives hope that such a
decomposition might exist. In contrast, a partial decomposition 
based on the Newman-Penrose formalism depends on the choice of a 
tetrad field, i.e., is 
gauge dependent even under `small' coordinate transformations \cite{baker}. Second, a conserved energy for the solutions of the wave 
equation exists, but the energy
density is negative inside the ergosphere. This fact 
excludes, at least a direct, application of 
the so called `energy methods' to a proof of stability of the solutions. The 
total energy could be finite while the field still might grow 
exponentially in parts of the 
spacetime. But recently a local stability result has been proved that 
the restrictions of 
the solutions to compact subsets $K$ in space are elements of 
$L^{\infty}_{\mathbb{C}}(K)$ with a norm converging to zero
for $t \rightarrow \infty$ \cite{finster}. Because of the 
absence of a decomposition of the Regge-Wheeler-Zerilli-Moncrief 
type, the question of  
applicability of the last and similar other results to the question 
of linearized stability of the Kerr metric is still open.  
\newline
\linebreak
As mentioned above, 
\cite{beyercraciun} contains the surprising find of a new 
symmetry operator that commutes with a normalized\footnote{In the following, the term `normalized' means multiplication from the left by 
the factor $1/g^{00}$ in Boyer-Lindquist coordinates. For instance, such a `normalization' takes place if the Klein-Gordon equation is solved for 
the second order time derivative of the unknown.}
form of the wave operator in a Kerr background.  Differently
to previously known symmetry operators for the wave operator, this 
operator contains only a partial time derivative of the first order, but not 
of higher order.  As a consequence, in formulations of the 
initial value problem for the wave equation 
in terms of first order systems of PDE and related formulations
such as \cite{beyer1}, this operator leads 
on a operator $\hat{S}$ 
that formally\footnote{i.e., on the level of tuples 
of functions that are differentiable 
to sufficiently high order.} commutes with the infinitesimal 
generator $G$ 
of time evolution and therefore is a candidate for the generator of a 
strongly continuous semigroup or 
group of symmetries. In precise terms, $\hat{S}$ 
should lead to an operator ${\cal S}$ that 
intertwines with the operators from the strongly continuous 
one-parameter group $T : [0,\infty) \rightarrow L(Y,Y)$
generated by $G$
\cite{beyer2}, \cite{kato5}, i.e., which is such that 
\begin{equation*}
{\cal S} \, T(t) \supset T(t) \, {\cal S}
\end{equation*}
for all $t \geq 0$. Here $Y$ denotes the space of data 
for the wave equation which is a complex Hilbert space 
\cite{beyer1}. The goal of the present note is to lay 
part of the foundation for a 
proof of the last in providing the new proof of the essential self-adjointness of the spatial 
part of the reduced normalized wave operator 
of the Kerr metric in a 
weighted 
$L^2$-space. 
In addition, it is believed that the employed method in this might be applicable 
to other separable wave equations. 
\newline
\linebreak
As a consequence, we also arrive at a purely operator theoretic proof of 
the well-posedness of the initial value problem of the reduced normalized
Klein-Gordon equation 
in the gravitational field of a Kerr black hole in the   
weighted  $L^2$-space. The last space was already used 
in \cite{beyer1}. Also the 
last reference
gives such a proof, but under the assumption that physical boundary conditions lead 
on the Friedrichs extension of the operator 
that is obtained 
from the spatial part of the 
reduced normalized wave operator of the Kerr metric
by restriction to $C^{\infty}$-functions with compact 
support. In particular, the results in this note 
prove that this assumption is justified.

%Uncomment out if preprint format required
%\pacs{00.00, 20.00, 42.10}

\section{The mathematical setting}

In Boyer-Lindquist coordinates\footnote{If not 
otherwise indicated, the symbols 
$t,r,\theta,\varphi$ denote coordinate projections whose 
domains will be obvious from the context. In addition, we assume 
the composition 
of maps, which includes addition, multiplication and so forth, 
always to be maximally defined. For instance, the sum of two 
complex-valued maps is defined on the intersection of their domains.
Finally, we use 
Planck units where 
the reduced Planck constant $\hbar$, the 
speed of light in vacuum $c$, 
and the gravitational constant $\gamma$, all have the numerical 
value $1$.},
$(t,r,\theta,\varphi) : 
\Omega \rightarrow {\mathbb{R}}^4$,
the Kerr metric $g$ is given by
\begin{align*}
& g = \left(1 - \frac{2Mr}{\Sigma}\right) dt \otimes dt + 
\frac{2Mar\sin^2\!\theta}{\Sigma} \, ( dt \otimes d\varphi 
+ d\varphi \otimes dt)
  \nonumber \\
& \qquad -
\frac{\Sigma}{\Delta} \, dr \otimes dr -
\Sigma \, d\theta \otimes d\theta  
- \frac{\Delta \overline{\Sigma}}{\Sigma} 
\sin^2\!\theta  \, d\varphi \otimes d\varphi \, \, , 
\end{align*}
where $M$ is the mass, $a \in [0,M]$ is the rotational
parameter and
\begin{align*} 
& \Delta := r^2 - 2Mr +a^2 \, \, , \, \, \Sigma := r^2 + a^2 
\cos^2\!\theta \, \, , \\
& 
\overline{\Sigma} := 
\frac{(r^2+a^2) \Sigma + 2 M a^2 r \sin^2\!\theta}{\Delta} 
= \frac{(r^2+a^2)^2}{\Delta} - a^2 \sin^2\!\theta \, \, , \\
& r_{+} := M + \sqrt{M^2-a^2}
\, \, , \, \, \Omega := {\mathbb{R}} \times (r_{+},\infty) \times
(-\pi,\pi) \times (0,\pi) \, \, .
\end{align*}
In these coordinates, the reduced wave equation, governing solutions
of the form 
\begin{equation*}
\psi(t,r,\theta,\phi) = \exp(im\varphi) \, u(t,r,\theta) \, \, ,
\end{equation*} 
where $m$ runs through all integers, is given by
\begin{align} \label{waveequation}
& \frac{\partial^2 u}{\partial t^2} + 
\frac{1}{\overline{\Sigma}} \cdot 
\left(i \, \frac{4mMar}{\triangle}\frac{\partial u}{\partial t}  - \frac{\partial}{\partial r} \triangle  \frac{\partial}{\partial r}
- \frac{m^2 a^2}{\triangle} - \frac{1}{\sin \theta} \,
\frac{\partial}{\partial \theta} \sin \theta \,\frac{\partial}{\partial \theta}
+ \frac{m^2}{\sin^2 \theta}\right) u = 0 \, \, .
\end{align}
The spatial part of the 
reduced normalized wave operator is given by 
\begin{align} \label{D2rtheta}
& D^2_{r \theta} f := \frac{1}{\overline{\Sigma}} 
\left( - \frac{\partial}{\partial r} \triangle  \frac{\partial}{\partial r}
- \frac{m^2 a^2}{\triangle} - \frac{1}{\sin \theta} \,
\frac{\partial}{\partial \theta} \sin \theta \,\frac{\partial}{\partial \theta}
+ \frac{m^2}{\sin^2 \theta}\right)f 
\end{align}
for every $f \in C^2(\Omega,{\Bbb{C}})$.
In particular, $D^2_{r \theta}$ 
is singular since the continuous extensions of the coefficients 
of its highest (second) order radial derivative vanish on the horizon 
$\{r_{+}\} \times [0,\pi]$. We represent the operator $D^2_{r \theta}$
as the operator $A$ below 
in the weighted $L^2$-space $X$ defined by   
\begin{equation*}
X := L_{C}^2\left(\Omega,g^{00} \sqrt{-|g|}\right) \, \, .
\end{equation*} 
Here $|g|$ denotes the determinant of the matrix $g_{ab}$. Note that
\begin{equation*}
g^{00} \sqrt{-|g|}  = 
\overline{\Sigma}
\sin \theta
\end{equation*}
is singular at the horizon. Hence the elements of 
$X$ vanish there in the mean. 
In the limit $a \rightarrow 0$, this weight reduces to the one 
that is commonly used in the
stability discussion of the Schwarzschild metric \cite{wald,kay}.

\section{Properties of the spatial part of the reduced normalized wave 
operator}

In the following, we prove that the operator $A_0$ defined below 
is linear, symmetric and essentially self-adjoint.\footnote{Note that 
the following operator  $A_0$ corresponds to the operator $A_{0}+C$ in 
\cite{beyer1}.}
\begin{definition}
\begin{itemize}
\item[]
\item[(i)]
We define the domain of $A_0$ to consist of 
all $f \in C^2(\bar{\Omega},{\mathbb{C}}) \cap X$ 
satisfying the conditions a), b) and c):    
\begin{itemize}
\item[a)] $D^2_{r \theta} f \in X$,
\item[b)]
there is $R > 0$ such that $f(r,\theta) = 0$ for all 
$r > R$ and $\theta \in I_{\theta} := (0,\pi)$, 
\item[c)]  
\begin{equation*}
\lim_{r \rightarrow r_{+}} 
\frac{\partial f}{\partial \theta}(r,\theta) = 0  
\end{equation*}
for all $\theta \in I_{\theta}$.
\end{itemize}
\item[(ii)] For every $f$ in the domain of $A_0$, 
we define 
\begin{equation*}
A_0 f := D^2_{r \theta} f \, \, . 
\end{equation*}
\end{itemize}
\end{definition}

\begin{lemma} $A_0$ is a densely-defined, linear and symmetric 
operator in $X$. In addition, $A_0$ is semibounded from 
below with lower bound 
\begin{equation*}
\alpha := - \frac{m^2 a^2}{4 M^2 r_{+}^2} \, \, .
\end{equation*}
\end{lemma}
\begin{proof}  In the following, 
$\braket{\quad|\quad}$ denotes the scalar product 
on $X$. Obviously, the domain of $A_0$ 
is a subspace of $X$ that contains $C^2_0(\Omega,{\mathbb{C}})$. 
Since the last is a dense subspace of $X$, that domain 
is a dense subspace of $X$. Further, $A_0$ is obviously
linear. In particular, it follows for $f,g$ from the domain of 
$A_0$ that 
\begin{align*}
& \overline{\Sigma} \sin \theta \, f^{*} A_0 g \\
& = 
\frac{\partial}{\partial r} \left\{
 \Delta \sin \theta 
\left[ \left(\frac{\partial f}{\partial r}\right)^{*} 
g - f^{*} \, \frac{\partial g}{\partial r} 
\right] \right\} + \frac{\partial}{\partial \theta} \left\{
\sin \theta 
\left[ \left(\frac{\partial f}{\partial \theta}\right)^{*} 
g - f^{*} \, \frac{\partial g}{\partial \theta} 
\right]
\right\} \\
& \quad \, \, + 
\overline{\Sigma} \sin \theta \, (A_0 f)^{*} g
\end{align*}
and hence by Green's theorem that 
\begin{align*}
& \braket{f|A_0 g} =
\int_{\Omega} \overline{\Sigma} \sin \theta \, f^{*} A_0 g
\, dr d\theta = 
\int_{\Omega} 
\overline{\Sigma} \sin \theta \, (A_0 f)^{*} g
\, dr d\theta \\
& = \braket{A_0 f | g} \, \, . 
\end{align*}
Hence, $A_0$ is symmetric. Further, it follows for $f$ from 
the domain of $A_0$ that
\begin{align*}
& \overline{\Sigma} \sin \theta \, f^{*} (A_0 - \alpha) f = 
- \frac{\partial}{\partial r} \left(
 \Delta \sin \theta  \, f^{*} \, \frac{\partial f}{\partial r} 
\right) - \frac{\partial}{\partial \theta} \left(
\sin \theta \, f^{*} \, \frac{\partial f}{\partial \theta} 
\right) \\
& + \sin \theta \left\{
\Delta \left| \frac{\partial f}{\partial r} \right|^2 + 
\left| \frac{\partial f}{\partial \theta} \right|^2 + \left(
- \alpha \overline{\Sigma} 
- \frac{m^2 a^2}{\Delta} + \frac{m^2}{\sin^2 \theta} 
\right) |f|^2
\right\} \\
& \geq - \frac{\partial}{\partial r} \left(
 \Delta \sin \theta  \, f^{*} \, \frac{\partial f}{\partial r} 
\right) - \frac{\partial}{\partial \theta} \left(
\sin \theta \, f^{*} \, \frac{\partial f}{\partial \theta} 
\right) \, \, ,
\end{align*} 
where it has been used that 
\begin{equation*}
\overline{\Sigma} \geq \frac{4 M^2 r_{+}^2}{\Delta} \, \, .
\end{equation*} 
Hence it follows by Green's theorem that 
\begin{align*}
& \braket{f|(A_0 - \alpha) f} = 
\int_{\Omega} \overline{\Sigma} \sin \theta \, f^{*} (A_0 
- \alpha) f
\, dr d\theta  \geq 0 
\end{align*}
and, finally, that $A_0$ is semibounded from 
below with lower bound $\alpha$.
\end{proof}

\begin{rem} We note that the domain of $A_0$ contains 
all products $f \otimes (P^{m}_{l} \circ \cos)$ where 
$f \in C^{2}_{0}(I_r,{\mathbb{C}})$
and $P^{m}_{l} : (-1,1) \rightarrow {\mathbb{R}}$
is the generalized Legendre polynomial corresponding 
to $m \in {\mathbb{Z}}$ and $l \in \{|m|, |m|+1, \dots\}$. 
\end{rem}

\begin{theorem}
$A_0$ is essentially self-adjoint. 
\end{theorem}

\begin{proof}
According to a well-known criterion for essential 
self-adjointness\footnote{E.g., see Section~X.1 in \cite{reedsimon}.}, 
it follows that $A_0$ is essentially self-adjoint if there is 
$\lambda < - \alpha$ such that the range of $A_0 - \lambda$
is dense in $X$. The existence of such $\lambda$ will be shown 
in the following. For this, we note that the elementary inequalities
\begin{equation} \label{basicineq}
\frac{r^4}{\triangle} \leqslant \overline{\Sigma}
\leqslant \frac{4M^2}{r_{+}^2} \,
\frac{r^4}{\triangle}  \, \, , \, \,
\end{equation}
imply that the underlying sets $X$ of 
$L_{\Bbb{C}}^2(\Omega, r^4 \sin \theta / \triangle)$ 
are equal and that the corresponding norms 
that are induced on that 
set are equivalent. These facts are basic for the following.
\newline
\linebreak
In a first step, we note that the following holds 
for arbitrary $\lambda \in {\mathbb{C}}$
and arbitrary $f$ in the domain of $A_0$ 
\begin{align} \label{fundrel}
& (A_0 -\lambda) f \\
& =
T_{r^4/(\Delta \overline{\Sigma})} 
\left[ 
\frac{1}{r^4/\Delta} 
\left( - \frac{\partial}{\partial r} \triangle  \frac{\partial}{\partial r}
- \frac{m^2 a^2}{\triangle} - \frac{1}{\sin \theta} \,
\frac{\partial}{\partial \theta} \sin \theta \,\frac{\partial}{\partial \theta}
+ \frac{m^2}{\sin^2 \theta} \right) - \lambda - 
\lambda \,  T_{(\Delta \overline{\Sigma}/r^4) - 1}
\right] f \nonumber 
\end{align}
where $T_{r^4/(\Delta \overline{\Sigma})}\,$, 
$T_{(\Delta \overline{\Sigma}/r^4) - 1}$ denote the maximal 
multiplication operators in $X$ with the functions 
$r^4/(\Delta \overline{\Sigma})$ and 
$(\Delta \overline{\Sigma}/r^4) - 1$, respectively. As a consequence 
of (\ref{basicineq}), $T_{r^4/(\Delta \overline{\Sigma})}$ is 
defined on the whole of $X$ 
as well as bounded and bijective. In addition, as a consequence 
of  
\begin{equation*}
0 \leq \frac{\Delta \overline{\Sigma}}{r^4}  - 1 \leq 5 \, \frac{a^2}{M^2}
\, \, ,
\end{equation*}
it follows that $T_{(\Delta \overline{\Sigma}/r^4) - 1}$ 
is a bounded positive self-adjoint operator on $X$. Motivated 
by (\ref{fundrel}), we define an auxiliary operator $H$
in $L_{\Bbb{C}}^2(\Omega, r^4 \sin \theta / \triangle)$ 
whose domain $D(H)$ coincides with the domain of $A_0$ and that is 
defined by 
\begin{align*}
H f := \frac{1}{r^4/\Delta} 
\left( - \frac{\partial}{\partial r} \triangle  \frac{\partial}{\partial r}
- \frac{m^2 a^2}{\triangle} - \frac{1}{\sin \theta} \,
\frac{\partial}{\partial \theta} \sin \theta \,\frac{\partial}{\partial \theta}
+ \frac{m^2}{\sin^2 \theta} \right) f \, \, .
\end{align*} 
for every $f \in D(H)$. Utilizing $H$, the identity 
(\ref{basicineq})
is equivalent to 
\begin{align} \label{fundrelnew}
& (A_0 -\lambda) f =
T_{r^4/(\Delta \overline{\Sigma})} 
\left( H - \lambda - 
\lambda \,  T_{(\Delta \overline{\Sigma}/r^4) - 1}
\right) f 
\end{align}
for every $f \in D(H)$. 
Note that, differently to 
$A_0$, $H$ can be obtained 
by `separation' from an operator which is in a certain sense 
`spherically symmetric'. This fact 
significantly simplifies the study of the properties of $H$. 
\newline
\linebreak
In the next 
step, we show that $H$ is a densely-defined, linear, symmetric
and essentially self-adjoint  
operator in $L_{\Bbb{C}}^2(\Omega, r^4 \sin \theta / \Delta)$ 
which is semibounded from 
below with lower bound 
\begin{equation*}
\beta := - \frac{m^2 a^2}{r_{+}^4} \, \, .
\end{equation*}
In this, $\braket{\quad|\quad}$ denotes the scalar product 
of $L_{\Bbb{C}}^2(\Omega, r^4 \sin \theta / \Delta)$. 
Indeed, it follows for $f,g \in D(H)$ that 
\begin{align*}
& \frac{r^4}{\Delta} 
\sin \theta \, f^{*} H g \\
& = 
\frac{\partial}{\partial r} \left\{
 \Delta \sin \theta 
\left[ \left(\frac{\partial f}{\partial r}\right)^{*} 
g - f^{*} \, \frac{\partial g}{\partial r} 
\right] \right\} + \frac{\partial}{\partial \theta} \left\{
\sin \theta 
\left[ \left(\frac{\partial f}{\partial \theta}\right)^{*} 
g - f^{*} \, \frac{\partial g}{\partial \theta} 
\right]
\right\} \\
& \quad \, \, + 
\frac{r^4}{\Delta} \sin \theta \,(H f)^{*} g
\end{align*}
and hence by Green's theorem that 
\begin{align*}
& \braket{f|H g} =
\int_{\Omega}  \frac{r^4}{\Delta} \sin \theta \, f^{*} H g
\, dr d\theta = 
\int_{\Omega} \frac{r^4}{\Delta} \sin \theta \,(H f)^{*} g
\, dr d\theta = \braket{H f | g} \, \, . 
\end{align*}
Hence, $H$ is symmetric. Further, it follows for $f \in D(H)$ 
that
\begin{align*}
& \frac{r^4}{\Delta}  \sin \theta \, f^{*} (H - \beta) f = 
- \frac{\partial}{\partial r} \left(
 \Delta \sin \theta \, f^{*} \, \frac{\partial f}{\partial r} 
\right) - \frac{\partial}{\partial \theta} \left(
\sin \theta \,f^{*} \, \frac{\partial f}{\partial \theta} 
\right) \\
& + \sin \theta \left\{
\Delta \left| \frac{\partial f}{\partial r} \right|^2 + 
\left| \frac{\partial f}{\partial \theta} \right|^2 + \left(
- \beta \, \frac{r^4}{\Delta}
- \frac{m^2 a^2}{\Delta} + \frac{m^2}{\sin^2 \theta} 
\right) |f|^2
\right\} \\
& \geq - \frac{\partial}{\partial r} \left(
 \Delta \sin \theta \, f^{*} \, \frac{\partial f}{\partial r} 
\right) - \frac{\partial}{\partial \theta} \left(
\sin \theta \,f^{*} \, \frac{\partial f}{\partial \theta} 
\right) \, \, .
\end{align*} 
Hence it follows by Green's theorem that 
\begin{align*}
& \braket{f|(H - \beta) f} = 
\int_{\Omega} \frac{r^4}{\Delta} \sin \theta \, f^{*} (H 
- \beta) f
\, dr d\theta  \geq 0 
\end{align*}
and, finally, that $H$ is semibounded from 
below with lower bound $\beta$. Hence it follows that 
$H$ is essentially self-adjoint if there is $\lambda < \beta$
such that the range of $H - \lambda$ is dense in 
$L_{\Bbb{C}}^2(\Omega, r^4 \sin \theta / \Delta)$. This fact 
will be proved by application of the theory of Sturm-Liouville 
operators. For this, we define $I_{r} := (r_{+}, \infty)$,  
\begin{equation*}
X_r := L_{\Bbb{C}}^2(I_{r},r^4/\triangle) 
\end{equation*}
and for every $l \in \{|m|,|m|+1,\dots\}$  
the Sturm-Liouville operators $A_{rml},A_{rml0}$ in $X_r$ by  
\begin{align*}
A_{rml0}f & := - \frac{1}{r^4/\Delta} \left(\triangle f^{\, \prime} \right)^{\, \prime} \\
A_{rml}f & := \frac{1}{r^4/\Delta} 
\left\{-\left(\triangle f^{\, \prime} \right)^{\, \prime} +  
\left[- \frac{m^2 a^2}{\triangle} + l(l+1) \right] f \right\}
\\
& \phantom{:}=
- \frac{1}{r^4/\Delta} \left(\triangle f^{\, \prime} \right)^{\, \prime} 
+ \frac{1}{r^4} 
\left[l(l+1) \, \Delta - m^2 a^2 \right] f 
\end{align*}
for every $f \in C^2_0(I_{r},{\Bbb{C}})$. Obviously, 
$A_{rml0}, A_{rml}$ are both densely-defined, linear and 
symmetric. The equation 
$\left(\triangle f^{\, \prime} \right)^{\, \prime}=0$
has nonvanishing constants as solutions. Since these are not in $X_{r}$
at both ends of $I_r$, it follows that $A_{rml0}$ is in the 
limit point case at $r_{+}$ and at $+\infty$. Hence $A_{rml0}$
is essentially self-adjoint (see, e.g., \cite{weidmann}). Further, 
since $[l(l+1) \, \Delta - m^2 a^2]/r^4$ is bounded and real-valued, 
it follows from that by the Rellich-Kato theorem, e.g., see Theorem X.12 
in Volume II of \cite{reedsimon}, that $A_{rml}$ is also essentially 
self-adjoint and that the domains of the closures of $A_{rml0}$ and 
$A_{rml}$, $\bar{A}_{rml0}$ 
and $\bar{A}_{rml}$, respectively, coincide. Obviously, 
$\bar{A}_{rml0}$ is semibounded from below with lower bound 
$\beta$. Hence it follows that the range, 
\begin{equation*}
\textrm{Ran}(A_{rml} - \lambda) \, \, ,
\end{equation*}
of $A_{rml} - \lambda$ is dense in $X_r$ for $\lambda < \beta$.
In the following, we assume that $\lambda < \beta$. 
We note that for every $f \in C^2_0(I_{r},{\Bbb{C}})$ and 
$l \in \{|m|,|m|+1,\dots\}$ 
\begin{equation} \label{basicrel}
(H - \lambda) [f \otimes (P_l^m \circ \cos)] = 
[(A_{rml} - \lambda) f] \otimes (P_l^m \circ \cos) \, \, .
\end{equation}  
Also, we denote the span of the elements of $D(H)$ of the form 
\begin{equation*}
f \otimes (P_l^m \circ \cos) \, \, , 
\end{equation*} 
where $f \in C^2_0(I_{r},{\Bbb{C}})$ and 
$l \in \{|m|,|m|+1,\dots\}$, by $D$. That $(H - \lambda)D$, 
and hence also $\textrm{Ran}(H-\lambda)$, is dense in $L_{\Bbb{C}}^2(\Omega, r^4 \sin \theta / \Delta)$ can be concluded as follows. 
For this, let $e_{0},e_1,\dots$ be some Hilbert basis 
of $X_r$. Since $P_{|m|}^{m} \circ \cos,P_{|m|+1}^{m} \circ \cos,
\dots$ is a Hilbert 
basis of $L_{\Bbb{C}}^2(I_{\theta},\sin)$, where $I_{\theta} := (0,\pi)$,
the family 
\begin{equation*}
\left(e_{k} \otimes f_{l} \right)_{(k,l) \in 
{\mathbb{N}} \times \{|m|,|m|+1,\dots\}}
\end{equation*}
is a Hilbert basis of 
$L_{\Bbb{C}}^2(\Omega, r^4 \sin \theta / \Delta)$, where 
\begin{equation*}
f_l := P_{l}^{m} \circ \cos
\end{equation*}
for every $l \in \{|m|,|m|+1,\dots\}$. Since 
$\textrm{Ran}(A_{rml} - \lambda)$ is dense in $X_r$, it follows 
by (\ref{basicrel}) that
\begin{equation*} 
e_{k} \otimes f_{l} \in \overline{(H - \lambda)D}
\end{equation*}
for every $(k,l) \in {\mathbb{N}} \times \{|m|,|m|+1,\dots\}$.
Since the span of the last family is dense in 
$L_{\Bbb{C}}^2(\Omega, r^4 \sin \theta / \Delta)$, this implies 
that $\overline{(H - \lambda)D}$ is dense in 
$L_{\Bbb{C}}^2(\Omega, r^4 \sin \theta / \Delta)$. Hence 
$H$ is also essentially self-adjoint. 
\newline
\linebreak
In the final step, we use (\ref{fundrelnew}) to prove that 
\begin{equation*}
\textrm{Ran}(A_0 - \lambda) 
\end{equation*}
is dense in $X$. Since $H - \lambda$ is essentially self-adjoint 
such that $\bar{H} - \lambda$ is bijective and since
\begin{equation*}  
- \lambda \,  T_{(\Delta \overline{\Sigma}/r^4) - 1} 
\end{equation*}
is a positive bounded self-adjoint operator, it follows by 
the Rellich-Kato theorem that densely-defined, linear and 
symmetric operator  
\begin{equation*}
H - \lambda - \lambda \,  T_{(\Delta \overline{\Sigma}/r^4) - 1}
\end{equation*}
is essentially self-adjoint and that the closure of this operator 
is bijective. Hence the range of this operator is dense in  
$L_{\Bbb{C}}^2(\Omega, r^4 \sin \theta / \Delta)$ as well as in $X$.
Finally, since  $T_{r^4/(\Delta \overline{\Sigma})}$ defines 
a bijective bounded linear operator in $X$, it follows by help
of (\ref{fundrelnew}) that 
\begin{equation*}
\textrm{Ran} (A_{0} - \lambda)
\end{equation*}
is dense in $X$. Since $\lambda < \beta \leq \alpha$, the last 
implies that $A_{0}$ is essentially self-adjoint.   
\end{proof}

\section{The case of the Klein-Gordon equation}
In the case of a Klein-Gordon field of mass $\mu \geq 0$, 
the equation corresponding to (\ref{waveequation})
is given by     
\begin{align} \label{kleingordon} 
& \frac{\partial^2 u}{\partial t^2} + 
\frac{1}{\overline{\Sigma}} \cdot 
\left(i \, \frac{4mMar}{\triangle}\frac{\partial u}{\partial t}  - \frac{\partial}{\partial r} \triangle  \frac{\partial}{\partial r}
- \frac{m^2 a^2}{\triangle} - \frac{1}{\sin \theta} \,
\frac{\partial}{\partial \theta} \sin \theta \,\frac{\partial}{\partial \theta}
+ \frac{m^2}{\sin^2 \theta} + \mu^2 \Sigma \right) u = 0 \, \, .
\end{align}
Hence in this case, the operator corresponding to 
$A_0$ is  
defined by 
\begin{equation*}
A_{0 \mu} := A_0 + T_h \, \, , 
\end{equation*} 
where $T_{h}$ denotes the maximal 
multiplication operator in $X$ by the real-valued function $h$
defined by 
\begin{equation*}
h := \mu^2 \, \frac{\Sigma}{\overline{\Sigma}} \, \, . 
\end{equation*} 
Since the last is also bounded, 
$T_{h}$ is a bounded self-adjoint
operator on $X$. Hence it follows by the 
Rellich-Kato theorem that $A_{0 \mu}$ is essentially self-adjoint 
if and only if $A_{0}$ is essentially self-adjoint. Since the last is 
the case, $A_{0 \mu}$ is essentially self-adjoint, too.

\section{Consequences}
As a consequence of the essential self-adjointness of 
$A_{0 \mu}$,
the objects $X, A_{\mu,-\alpha+\varepsilon} := A_{\mu} 
- \alpha + \varepsilon,B$ and 
$C : - (- \alpha + \varepsilon)$ are easily seen 
to satisfy Assumptions~$1$ 
and $4$ of \cite{beyer}.\footnote{See also the Section~$5.1$ 
on `Damped wave equations' in \cite{beyer2}.} Here $A_{\mu}$
denotes the closure of $A_{\mu 0}$ and
$\varepsilon > 0$
is assumed to have the dimension $l^{-2}$. The exact value of 
$\varepsilon$ does not influence the results in any essential way. 
In addition, $B$ denotes the maximal multiplication operator
in $X$ by the function multiplying $i \, \partial u / \partial t$
in (\ref{kleingordon}). Since that function is bounded and positive 
real-valued, $B$ is a bounded linear and positive self-adjoint operator
on $X$ given by 
\begin{equation*}
Bf =  \frac{4mMar}{\triangle \overline{\Sigma}} \, f
\end{equation*}
for every $f \in X$.
Hence, application of the results of \cite{beyer} 
give, in particular, the following well-posed formulation of the 
initial value problem for (\ref{kleingordon}).
\begin{theorem}
\begin{itemize}
\item[]
\item[(i)] By
\begin{equation*} 
Y := D(A_{\mu,-\alpha+\varepsilon}^{1/2}) \times X 
\end{equation*}
and 
\begin{equation*}
(\xi | \eta) := \braket{A_{\mu,-\alpha+\varepsilon}^{1/2}\,\xi_1|
A_{\mu,-\alpha+\varepsilon}^{1/2}\,\eta_1} + \braket{\xi_2|\eta_2}
\end{equation*}
for all $ \xi = (\xi_1,\xi_2), \eta = (\eta_1,\eta_2) \in Y$,
there is defined a complex Hilbert space $(Y,(\, |\, ))$. 
\item[(ii)] The operators
$G$ and $-G$ defined by 
\begin{equation*}
G(\xi,\eta) := (-\eta, A_{\mu} \, \xi + iB\eta) \, \, 
\end{equation*}
for all $\xi \in D(A_{\mu})$ and $\eta \in 
D(A_{\mu,-\alpha+\varepsilon}^{1/2})$
are infinitesimal generators of strongly continuous semigroups
$T_{+} : [0,\infty) \rightarrow L(Y,Y)$ and 
$T_{-} : [0,\infty) \rightarrow L(Y,Y)$, respectively. 
\item[(iii)]
For every $t_0 \in {\Bbb{R}}$ and every $\xi \in  
D(A_{\mu}) \times D(A_{\mu,-\alpha+\varepsilon}^{1/2})$, 
there is a uniquely determined 
differentiable map $u : {\Bbb{R}} \rightarrow Y$
such that 
\begin{equation*}
u(t_0) = \xi
\end{equation*}
and 
\begin{equation*} 
u^{\prime}(t) = - G u(t) 
\end{equation*}
for all $t \in {\Bbb{R}}$. Here $\, ^{\prime}$ denotes differentiation 
of functions assuming values in $Y$. Moreover, this
$u$ is given by 
\begin{equation*}
u(t) :=     
 \left\{
 \begin{array}{cl}
 T_{+}(t)\xi & \text{for $t \geqslant 0$} \\
 T_{-}(-t)\xi & \text{for $t < 0$}
 \end{array}
 \right. 
\end{equation*}
for all $t \in {\Bbb{R}}$.
\end{itemize}
\end{theorem}

\subsection*{Acknowledgments}
 
The author is thankful for the hospitality and support
by the Sections `Geometric Analysis and Gravitation' (G. Huisken)
and `Astrophysical Relativity' (B. F. Schutz)  
of Max-Planck-Institute for Gravitational Physics (AEI)
and the Institute for Theoretical Astrophysics (TAT,  
K. Kokkotas)  
at the University of Tuebingen. This work was supported by the 
German Foundation for 
Research (DFG) via the SFB/TR7 grant.

\section{Appendix}

\end{document}